\documentclass[aip,jap,reprint,floatfix]{revtex4-1}

\usepackage{amsmath}
\usepackage{amssymb}
\usepackage[usenames,dvipsnames,svgnames,table]{xcolor}
\usepackage{graphicx}
\usepackage[normalem]{ulem}
\usepackage{hyperref}
\hypersetup{
    colorlinks,
    linkcolor={blue},
    citecolor={blue},
    urlcolor={blue}
}
\usepackage[T1]{fontenc}
\usepackage{newtxtext,newtxmath}
\usepackage{mathtools}
\usepackage[tracking=true,protrusion=true,expansion=true]{microtype}
\usepackage{bm}
\usepackage{siunitx}
\usepackage{xfrac}

\newcommand{\vect}[1]{\bm{#1}} 
\newcommand{\matr}[1]{\bm{#1}} 

\begin{document}


\title{Modeling shape selection of buckled dielectric elastomers}



\author{Jacob Langham}
\email[]{J.Langham@bristol.ac.uk}
\affiliation{School of Mathematics, University of Bristol, Bristol, UK, BS8 1TW}
\affiliation{Mathematics Institute, University of Warwick, Coventry, UK, CV4 7AL}
\author{Hadrien Bense}
\email[]{hadrien.bense@espci.fr}
\affiliation{Laboratoire de Physique et M\'{e}canique des Milieux
H\'{e}t\'{e}rog\`{e}nes (PMMH), CNRS, ESPCI Paris, \\ PSL Research University, Sorbonne Universit\'{e}, Univ.\ Paris Diderot, Paris, France}
\author{Dwight Barkley}
\email[]{D.Barkley@warwick.ac.uk}
\affiliation{Mathematics Institute, University of Warwick, Coventry, UK, CV4 7AL}

\date{\today}

\begin{abstract}
\rightskip1in
A dielectric elastomer whose edges are held fixed will buckle, given sufficient
applied voltage, resulting in a nontrivial out-of-plane deformation. We study
this situation numerically using a nonlinear elastic model which decouples two
of the principal electrostatic stresses acting on an elastomer: normal pressure
due to the mutual attraction of oppositely charged electrodes and tangential
shear (``fringing'') due to repulsion of like charges at electrode edges. 
These
enter via physically simplified boundary conditions that are
applied in a fixed reference domain using a nondimensional approach.
The method is valid for small to moderate strains and is
straightforward to implement in a generic nonlinear elasticity code.
We validate the model by directly comparing simulated equilibrium shapes with
experiment. For circular electrodes which buckle axisymetrically, the shape of
the deflection profile is captured. Annular electrodes of different widths
produce azimuthal ripples with wavelengths that match our simulations. 
In this case, it is essential to compute multiple equilibria because the first
model solution obtained by the nonlinear solver (Newton's method) is often not
the energetically favored state. We address this using a numerical technique
known as ``deflation''. 
Finally, we observe the large number of different solutions that may be obtained
for the case of a long rectangular strip.  
\end{abstract}

\maketitle 
%
%
\section{Introduction}
Dielectric elastomers (DEs) are a class of soft and flexible actuating devices
that deform when subjected to electric fields.  The significant mechanical
strains available in DE systems, compared with competitive technologies, has
driven their development in numerous contexts, particularly in
engineering.\cite{OHalloran2008}  In a number of applications, including
pumps,\cite{Pelrine2001,Takavol2014,Takavol2016}
loudspeakers,\cite{Heydt2000,Heydt2006} tactile displays\cite{Carpi2010,
Vishniakou2013} and others~\cite{Jung2007,Son2012} a key component is a
purposefully induced buckling instability. 

Typical DE setups involve a thin elastomer membrane coated on opposite faces
with areas of conducting material, thereby partitioning the surface into
electrically ``active'' and ``inactive'' regions.  A connecting circuit turns
the active regions into oppositely charged electrodes. This creates a flexible
capacitor in which the intervening dielectric (the elastomer) is apt to deform
under the influence of electrostatic forces.  The conducting material is
fabricated so that it is free to bend and stretch with the elastomer without
constraining its movement.
%

\begin{figure}[ht]
	\centering%
  \includegraphics[width=8.5cm]{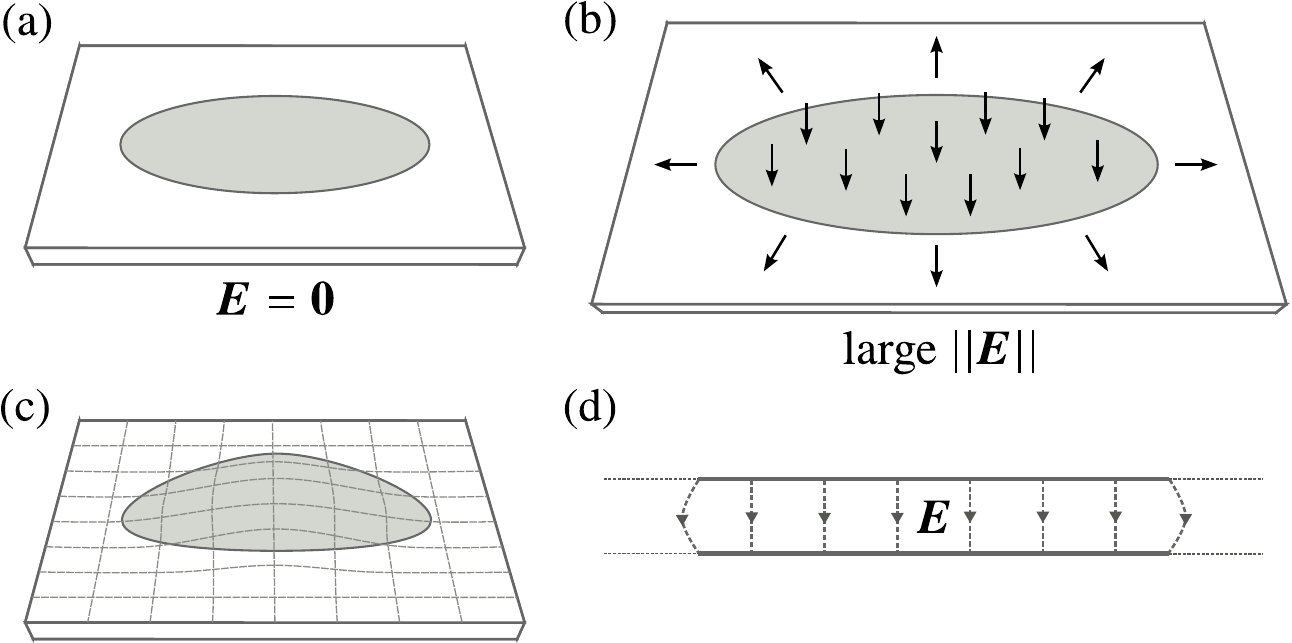}%
    \caption[Diagrams explaining the operation of a dielectric elastomer.]
    {
	Diagrams of a dielectric elastomer in different situations. In
        parts (a)--(c) the
        top electrode (active region) is shaded in gray. 
	(a)~DE with no applied electric field
	($\vect{E} = \vect{0}$). 
	(b)~DE with a strong applied electric
	field (large $||\vect{E}||$),
	which causes the elastomer to deform. If
	the boundaries of the medium are free to move, the material compresses
	in the thickness direction and extends laterally. 
	(c)~If the edges of the medium are instead
	held fixed, the electrostatic forces between the electrodes force the
	elastomer to buckle out-of-plane.  
        (d)~Two-dimensional cross-section
        through the middle of the elastomer showing the fringing of the electric
		field $\vect{E}$. The top and bottom electrodes are represented by thick gray
        lines. Dashed lines with arrows indicate the direction of the electric
        field. At the center the field lines are uniformly spaced and normal to the
        electrodes.  At the edges they warp, leaving a small nonzero component of
        electric field tangent to the medium surface.}%
	\label{fig:elastomerSchematics}
\end{figure}

%
Figure~\ref{fig:elastomerSchematics}(a) shows an example DE geometry in its zero
strain configuration, before any electric field $\vect{E}$ has been applied.
The medium is a thin cuboid, on which the top electrode can be seen, shaded in
gray. 
%
%
Typical materials used in applications are isotropic and incompressible. They
produce significant strains in response to applied voltages on the order of
kilovolts.  Figure~\ref{fig:elastomerSchematics}(b) demonstrates the effect of
an applied electric field, in a simple situation in which the lateral sides of
the medium are unconstrained.  When the voltage is turned on, attractive forces
arising from the charge imbalance on the two electrodes push the top and bottom
faces of the elastomer together. This compression is coupled to lateral
expansion of the film via incompressibility.
If instead, the edges of the elastomer are held fixed in
space, the active region and surrounding area will buckle out-of-plane as shown
in Fig.~\ref{fig:elastomerSchematics}(c). 
This is an inevitable consequence of the incompressible material
preserving its volume under compression of the electrodes.
The equilibrium shape adopted by a deformed elastomer is frequently
nontrivial and can contain waves or
wrinkles.\cite{Pelrine2000,Plante2006,Kollosche2012,Zhu2012,Bense2017}

Figure~\ref{fig:elastomerSchematics}(d) sketches the electric field lines
between the active regions in the ideal undeformed setting. For the most part,
the field is constant between the electrodes and electrostatic forces act
perpendicular to the top and bottom surfaces of the elastomer. At the conductor
edges the field lines become slightly curved because mutual repulsion of like
surface charges does not balance, as it does in the center. The result is a
fringing field with a small nonzero component tangent to the elastomer surface.
This picture also holds approximately for DEs after buckling due to the
small interstitial length scale. 
It should be noted that the electric field schematic is based on an idealized
understanding of a classical parallel-plate capacitor and may not always reflect
the experimental situation in a DE, where the charge distribution may not be
completely uniform.  Nevertheless, it suits our purposes here, as will become
clear.

The aim of this paper will be to numerically model buckled DE shapes and make
direct comparisons with experimental deformation profiles and images.  We
propose a straightforward approach to DE modeling based on a significant
simplification of the underlying physics (including the fringing effect), which
is nonetheless able to match nontrivial buckling shapes. 
%
%
\section{Experiment}
\label{sec:experiment}
The DEs used in the experiments are made of polyvinyl siloxane (PVS) with a
Young's modulus of \SI{250+-15}{\kPa}, estimated with a standard tensile test on
a strip.  After mixing equal amounts of base and catalyst, the liquid PVS is
spincoated at 500\,rpm for 15 seconds and cured, obtaining a solid disc of
approximate thickness \SI{150}{\um}.
The electrodes are made of carbon black powder brushed onto the top and bottom
surfaces of the cured polymer using a stencil.
They do not change the mechanical properties of the surface. Moreover, the
adhesion of the powder is remarkably good and the conductivity of the surface is
maintained across the full range of strains that we achieve.  The resistivity of
the coating is in the order of a few hundred kilohms.  

\begin{figure}[htbp]
	\centering%
  \includegraphics[width=0.8\columnwidth]{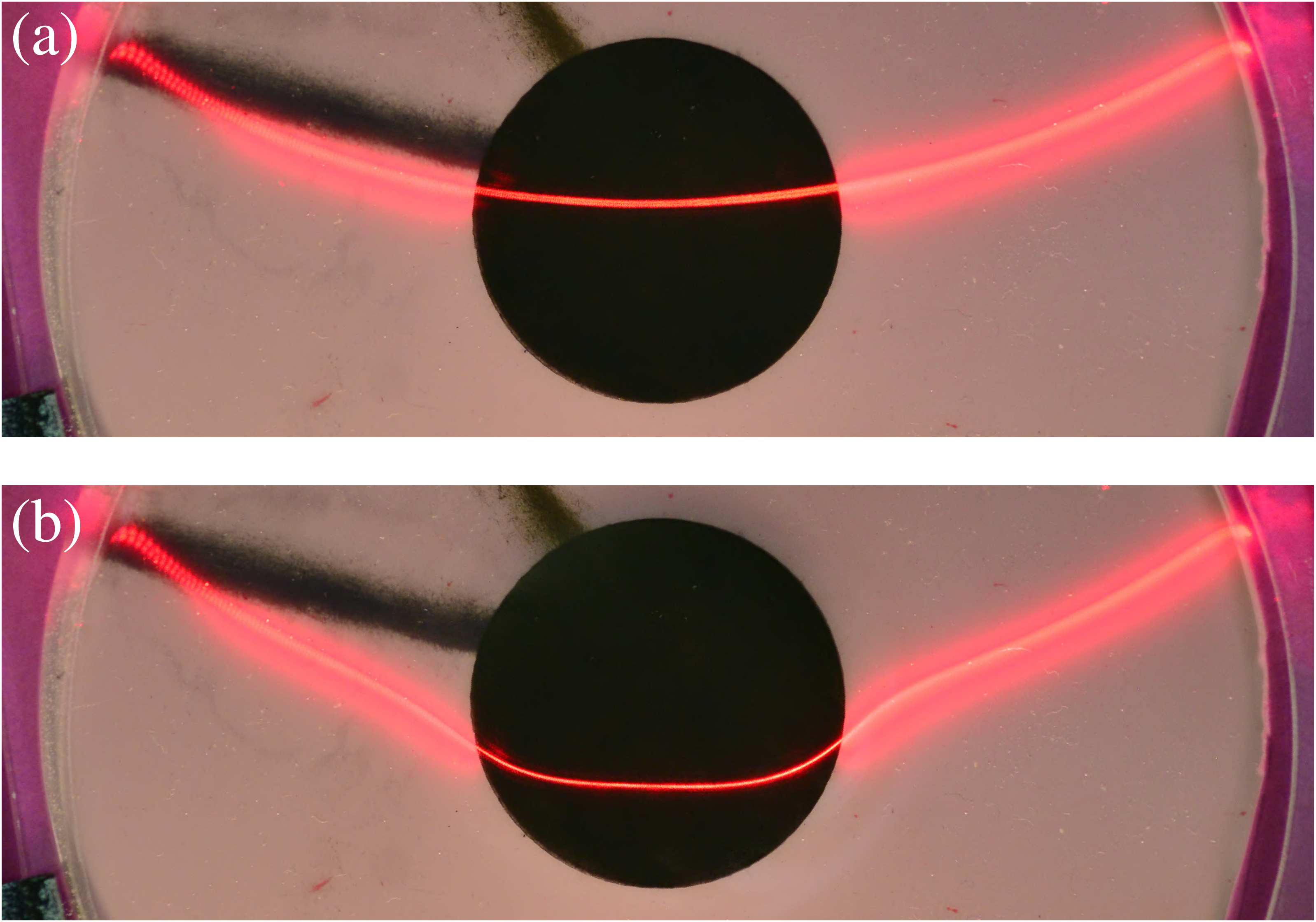}%
	\caption[The dielectric elastomer experiment with circular active region,
    before and after actuation.]{
	Overhead view of the experiment with circular active region of diameter 
		\SI{30}{\mm}, before and after
		actuation.  The applied voltages
		are: (a) \SI{0}{\kV} and (b) \SI{4}{\kV}.
	}%
	\label{fig:experiment_disc}
\end{figure}

Figure~\ref{fig:experiment_disc} shows the experiment in use.
The
DE is clamped in a rigid circular polyvinyl chloride (PVC) frame, of
diameter \SI{10}{\cm} without prestretching the material.
In the absence of applied voltage it sags under its own weight. 
A laser sheet is projected across
the diameter of the active region at an oblique incidence angle.
Its deflection---monitored using a camera directly above the experiment---is
proportional to the deflection of the elastomer surface,
allowing us to measure the vertical deformation of the system.  At the beginning
of the experiment, as the voltage is increased from zero, the membrane sags more
and more as the central region grows. However, after a threshold voltage is
reached, the system buckles, undergoing an axisymmetric deformation that is
strongly localized in the active region. In this paper, we are concerned with
capturing deformations after this initial buckling instability, for a variety of
active region shapes.
For DEs with circular electrodes at still higher voltages, a 
secondary instability has been observed that causes azimuthal wrinkling at
the electrode edges.\cite{Bense2017}

%
\section{Model}

Anticipating the nontrivial deformations observed in applications, we place
our elastomers in a nonlinear elasticity setting.  At equilibrium, they obey
the elastostatics equation
\begin{equation}
    \nabla \cdot \matr{\sigma}(\vect{x}) + \vect{b}(\vect{x}) = \vect{0},
    \label{eq:elastostatics}
\end{equation}
where $\matr{\sigma}$ is the Cauchy stress tensor, $\vect{b}$ is a body force
(density) and the equation is posed over all of the material points $\vect{x}$
that comprise the deformed object.  By specifying both an appropriate
constitutive law and boundary conditions this equation can be solved for the
deformation of the elastic body. In our model, electrostatic forces enter
the system via prescribed traction boundary conditions, which we shall
detail shortly.

Equation~\eqref{eq:elastostatics} is closed by specifying a
particular strain energy density function $W$. We shall use
the isotropic Mooney-Rivlin constitutive law, which in its incompressible formulation is
\begin{equation}
        W = c_1 (I_1 - 3) + c_2 (I_2 - 3),
        \label{eq:mooneyRivlin}
\end{equation}
for phenomenological model parameters $c_1, c_2$, where $I_1$ and $I_2$ denote
the first and second principal invariants of the Cauchy-Green strain tensor.
A variety of more sophisticated laws, including the Ogden, Gent, Yeoh and
Arruda-Boyce models, have been used in prior DE modeling studies. These
capture
elastomer strain responses with greater accuracy, especially at large
strains.\cite{Goulbourne2005,Wissler2005,Wissler2007a,Xu2010,Li2013}  However,
no prestretch is applied to the elastomers in our experiments and we consider
only moderate strains. In this regime we find the Mooney-Rivlin law to be more
than adequate for our purposes.
Moreover, an advantage to this model is that it only depends on two parameters:
$c_1$ and $c_2$. While elastomers can exhibit viscoelastic
properties,\cite{Sommer-Larsen2004,Plante2007} we shall work in the static
setting only and therefore need not consider viscoelasticity here.

We account for the effect of gravity with a constant body force density
$\vect{b}$ that acts vertically downwards. It has magnitude $\rho g$, where
$\rho$ is the material density (assumed to be constant) and $g$ is
gravitational acceleration. This is the only body force that appears in the
model. 

Finally and most importantly, we model the electrostatic forces present 
in the system, due to the surface charge distributions on the electrodes.
These dictate components of the Cauchy stress across the surfaces of the elastic
medium and hence the boundary conditions for Eq.~\eqref{eq:elastostatics}.
Specifically, if an (area) force density $\vect{\tau}$ impinges on the surface
of the deformed body with unit normal $\vect{n}$, then $\matr{\sigma}\vect{n} =
\vect{\tau}$. The \emph{traction} vector $\vect{\tau}$ is determined by
modeling considerations.

The principal traction is due to attractive forces between the oppositely
charged electrodes. The electrostatic pressure (force density) $p$ between
charged surfaces held at a voltage $V$ and at a separation distance $D$ is
\begin{equation}
    p = -\frac{1}{2}\epsilon \left( \frac{V}{D} \right)^{\!2}\!\!,
    \label{eq:classicPressure}
\end{equation}
where $\epsilon$ is the permittivity of the region between the charges. This
dictates the 
Cauchy stress acting in the surface normal direction within the active regions.

Towards the edges of the active regions, the electrostatic force on the charge
distribution has an additional component that is tangent to the surface of the
elastomer film. 
This arises because the repulsive forces between like charges
there are not balanced, as they are in the center. This is indicated in
Fig.~\ref{fig:elastomerSchematics}(d), which shows the resultant fringing of
the electric field lines at the edges.
The tangential forces are small, compared with the normal attraction between the
electrodes.
Nevertheless, they cause the compliant electrodes to stretch and pull on the
material to a certain extent.  

\begin{figure}[htbp]
    \centering%
    \includegraphics[width=8.5cm]{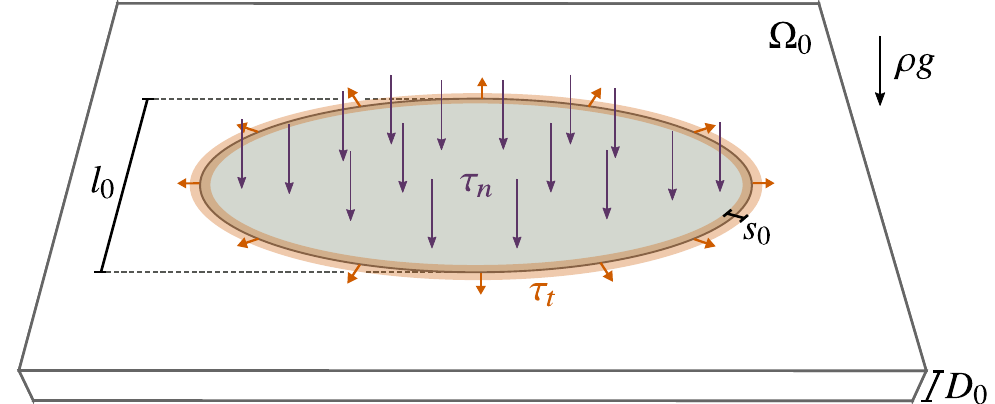}%
    \caption{Schematic of the model reference
              configuration $\Omega_0$, showing the application of normal
              pressure $\tau_n$ (purple arrows), tangential traction $\tau_t$
              (orange arrows), body force density $\rho g$ and the important
              length scales $D_0$, $l_0$ and $s_0$.}%
    \label{fig:modelSchematic}
\end{figure}

In this work, we make the simplifying assumption that the above surface forces
can be effectively captured by two regions of \emph{constant} traction
corresponding to the normal pressure and tangential fringing effect.
A schematic of our approach is shown in Fig.~\ref{fig:modelSchematic}.

While the electrostatic forces are physically manifest on the surfaces of the
deformed body, in practice our numerical simulations use a standard Lagrangian
coordinate system corresponding to a fixed, zero-strain reference domain
$\Omega_0$, as seen in the figure. Hence, Eq.~\eqref{eq:elastostatics} and
its boundary conditions must be referred back to this configuration.
Specifically, we solve
\begin{equation}
\nabla \cdot \matr{S}(\vect{X}) + \vect{b}_0(\vect{X}) = \vect{0},
    \label{eq:elastostaticsRef}
\end{equation}
%
where $\matr{S}$ is the (first) Piola-Kirchhoff stress tensor and
$\vect{b}_0$ is the body force. These quantities are $\matr{\sigma}$ and
$\vect{b}$ respectively, written in the Lagrangian frame and are defined over
all reference points $\vect{X} \in \Omega_0$. Spatial derivatives in
Eq.~\eqref{eq:elastostaticsRef} are taken with respect to the reference
co-ordinates. The traction boundary condition in this setting is
$\matr{S}\vect{N} = \vect{\tau}_0$, where $\vect{N}$ is the unit normal vector
field on the surface of $\Omega_0$. 
	Specification of the reference traction vector $\vect{\tau_0}$ is the point
	at which electrostatic forces enter our model.

The normal electrostatic pressure is set by a constant traction at
the top and bottom electrode surfaces of magnitude $\tau_n$, directed into the
reference material body.  The tangential fringing effect is applied in a small
annular neighborhood of width $s_0$, along the active region perimeter (see
Fig.~\ref{fig:modelSchematic}). Its magnitude $\tau_t$ is constant across this
region and its direction is given by the outward normal to the boundary of the
annulus.  
The sum of these two orthogonal vectors at each point comprises the model
reference traction $\vect{\tau_0}$.
It is important to note that in the physical system, the two traction
directions lie perpendicular and tangent to the surface of the deformed
configuration. By posing them in the Lagrangian frame, we introduce a
computationally convenient assumption that is only reasonable when strains are
not too large. The values of $\tau_n$ and $\tau_t$ will be discussed
momentarily. 

In addition to the width $s_0$, there are two important length scales present in
the model: the thickness of the unstrained domain $D_0$ and the characteristic
length of the active region $l_0$. The exact definition of $l_0$ depends on the
shape of the particular active region. In the case of the circle, it refers to
the diameter.

For a particular DE configuration, any ratio of electrostatic force
components will not change if the potential difference across the plates is
altered. This is simply because the electrostatic equations are linear.
%
%
Therefore, the ratio $\tau_t / \tau_n$ seems like a natural candidate for a
dimensionless parameter that determines the relative strength of the
tangential traction applied in the model. However, a better choice is
\begin{equation}
\kappa \coloneqq
\frac{s_0}{D_0}\frac{\tau_t}{\tau_n},
\label{eq:kappa}
\end{equation}
which takes into account the length scales of the problem. To see why this is
necessary, let us consider a potential difference $V$ between the active
surfaces in the undeformed geometry of Fig.~\ref{fig:modelSchematic}.
We know from Eq.~\eqref{eq:classicPressure} that the normal pressure on each
surface scales with $(V/D_0)^2$. 
Likewise, it may be shown (e.g.\ using the Maxwell stress tensor) that the
fringing force at the active region edges scales with $V^2/D_0$.
%
%
The corresponding quantity in our model is $s_0\tau_t$.
Therefore, electrostatics implies
that $s_0\tau_t/D_0\tau_n$ should be constant
with respect to changes in $D_0$.  
This scaling might cease to hold in cases where $D_0$ becomes comparable
to $l_0$, but in all cases we consider $D_0 \ll l_0$.

In the system with deformations, the fringing
force $f_t$ and normal pressure $p$ will be such that $f_t / Dp$ is constant,
where $D$ the deformed thickness.  To keep this constant in our model, it would
be strictly necessary to apply a correction, allowing both $\tau_t/\tau_n$ and
the directions of the applied tractions to vary with the deformed geometry.
Whilst we have investigated such an approach, it is fundamentally more
complicated and does not appear to be any more predictive for the phenomena
considered in this study.  Hence, we have opted for the simplicity of
maintaining constant $\kappa$, as defined in Eq.~\eqref{eq:kappa}.


Through the dimensionless parameter $\kappa$, we dictate the relative 
strength of the tangential fringing force applied in the model in a
geometry-independent way. Note that $\kappa = 0$ means no tangential 
shear and that larger $\kappa$ corresponds to a larger relative strength of
$\tau_t$.
%
%
The value of $\kappa$ is investigated in Sec.~\ref{sec:results}. 
An implicit, but reasonable assumption in defining $\kappa$ the way we
do is that solutions to the model system are not significantly affected by the
width $s_0$, provided that $s_0$ is sufficiently small relative to $D_0$
and $l_0$. This was verified in detail for
the \SI{3}{\kV} result presented later in Fig.~\ref{fig:fitmultiplevoltages}.  
In practice, we observe that for the thin simulation domains considered
	herein, setting $s_0$ smaller than $l_0$ is all that is essential.
	Indeed, it was
necessary for such geometries that $s_0$ be comparable to $D_0$ in order to ensure
that $s_0$
covered a sufficient number of points in the spatial discretization
of Eq.~\eqref{eq:elastostaticsRef}.

The above treatment is a deliberately straightforward and practical attempt to
access some of the shapes adopted by buckling DEs. It is worth reiterating here
that while it is physically motivated, our model is a simplification
of the full physics.
The complete picture is very complicated, since it involves a spatially-varying
charge distribution whose equilibrium configuration is coupled to the mechanical
deformation. Numerous prior studies have therefore opted to solve electrostatic
equations and an elasticity model (or viscoelasticity model) in
concert.\cite{Gao2011,Park2012,Vertechy2012,Khan2013,Klinkel2013,Park2013a,Henann2013,Vogel2014,Seifi2016,Wang2016}
Further detail may be added to the physical picture by accounting for complex
interactions arising from polarization of the dielectric and strain-dependent
permittivity.\cite{Zhao2007,Zhao2008a}

Of particular relevance to our study is the work of Vertechy \emph{et
al.}~\cite{Vertechy2012} who considered `diaphragm actuators'---buckled circular electrodes within a
rigid inactive region. By solving for the electric field both inside the DE and
in the surrounding free space, fully coupled with the elastostatic problem, they
were able to accurately match experimentally observed displacements. 
Also notable is the recent observation by Wang \emph{et al.}~\cite{Wang2016} of
a (simulated) instability in a diaphragm actuator, similar in character
to both the secondary instability of Ref.~[\onlinecite{Bense2017}] and the wavy patterns
that we demonstrate below for annular electrodes.
%

A simpler modeling approach, derived from the field theory of Suo \emph{et
al.}~\cite{Suo2008} treats the electric field in the Lagrangian frame as
constant and perpendicular to the electrodes, its effect on the mechanical
stress mediated via a free-energy function defined throughout the material. This
level of detail can be sufficient to capture many out-of-plane deformations and
instabilities well.\cite{Zhou2008,Zhao2008}
Another way to simplify matters (at least computationally speaking) is to reduce
the underlying equations to two spatial dimensions. This was used in
Ref.~[\onlinecite{OBrien2009}] to model DEs attached to frames that bend and curl when
activated.

By making the various simplifications detailed above, we sacrifice a certain
degree of precision in favor of a more conceptually straightforward model.  We
argue that there are only two electrostatic effects of principle importance: the
normal pressure and the fringing traction. Moreover, we are content to treat
these in a fixed reference frame, independent of medium deformation.
%
%
When applying our model, we use a nondimensional approach, explained in
Sec.~\ref{sec:results}.
%
This means that we need not worry about matching the effective pressure with the
exact voltage and deformed material thickness. Instead, model parameters are
fitted such that the applied tractions scale in a manner consistent
with Eq.~\eqref{eq:classicPressure}.
\section{Methods}
\label{sec:methods}
We perform nonlinear elasticity simulations using the finite element continuum
mechanics solvers from the \texttt{Chaste} software libraries,\cite{Chaste2013}
which provide an incompressible nonlinear elasticity
implementation that we modified for our own
purposes. 
The nonlinear solver is a damped Newton's method and the linear
solver is GMRES with \texttt{PETSc}'s additive
Schwarz preconditioner, using LU factorization blocks.\cite{Petsc}
The deformation map is solved on a zero-strain reference domain
$\Omega_0$, as depicted in Fig.~\ref{fig:modelSchematic}, using
tetrahedral quadratic elements. Meshes are constructed using
\texttt{Gmsh},\cite{Geuzaine2009} with a minimum of two layers of
tetrahedra in the thickness direction. To reduce the number of degrees of
freedom these are refined more at the active region and towards the center
where most of the strain occurs.  Furthermore, we allow elements in the
reference domain to be longer in the transverse direction than they are in
their thickness. The ratio of these respective dimensions is approximately
$1.5:1$ near the active regions and $10:1$ by the outer Dirichlet boundaries
where there is very little deformation.  In spite of these optimizations,
the aspect ratios of the physical system dictate that even the coarsest
possible meshes have many elements---typically our simulations use
on the order of $10^5$ degrees of freedom.

\subsection{Multiple solutions}
\label{sec:deflation}
The elastostatics equation [Eq.~\eqref{eq:elastostatics}] can have multiple
solutions. Consequently, there may be many different shapes that an elastomer
can adopt in which the material is in equilibrium with the external forces
imposed on it. This presents us with a problem when attempting to predict the
shape of a DE: the solution that nature selects may not be the one that we
happen to find using our nonlinear solver. To address this issue, we 
implemented an algorithm called ``deflation'', whose use in the context of
numerical PDE solving is due to Farrell \emph{et al.}\cite{Farrell2015}

The basic idea behind deflation is to factor out solutions from a PDE system
that are already known. In our case, we seek the zeros of a nonlinear operator
$\mathcal{F}$ defined by $\mathcal{F}(\vect{u}) = \nabla \cdot
\matr{S}(\vect{u}) + \vect{b}_0$, subject to the boundary conditions of our
model. Suppose that we have found solutions 
$\vect{u}^1,\ldots,\vect{u}^n$ already. Then we solve the deflated system
\begin{equation}
	\mathcal{G}(\vect{u}; \vect{u}^1, \ldots, \vect{u}^n) 
	\coloneqq 
	\left( \alpha + \sum_{i=1}^n
	\frac{1}{||\vect{u}-\vect{u}^i||^q} \right) \mathcal{F}(\vect{u}) = \vect{0}
	\label{eq:deflationOperatorBetter}
\end{equation}
for some $\alpha, q > 0$.
The deflated system
has the same solution set as $\mathcal{F}(\vect{u})
= \vect{0}$, less the known solutions
$\vect{u}^1,\ldots,\vect{u}^n$.
Any solution that we find to $\mathcal{G}(\vect{u}) = \vect{0}$
is therefore a new equilibrium state 
for the DE. 
The inclusion of the parameter $\alpha$ dissuades the numerical method from
improperly minimizing the residual of $\mathcal{G}$ 
below the solver tolerance by
pushing intermediate guesses further and further from the known solutions.

The procedure to solve Eq.~\eqref{eq:deflationOperatorBetter} was implemented using
\texttt{PETSc}.\cite{Petsc}
The augmentation of the nonlinear operator 
results in a rank-one update to the system Jacobian, causing it to lose its
sparsity. Consequently, whenever it is needed its application is performed in
terms of the Jacobian of the original system via matrix-free methods. Similarly,
the preconditioner is implemented matrix-free and is computed via the original
preconditioner using the Sherman-Morrison formula as suggested
in Ref.~[\onlinecite{Farrell2015}]. Below, we give some practical details concerning how
deflation was used to find multiple DE shapes.

Controlling the order of the singularities in the deflated system with $q$
affects how close any additional candidate solutions can get to $\vect{u}^1,
\ldots, \vect{u}^n$, as does varying $\alpha$.  Selection of these parameters
can greatly alter which solutions can be found by the nonlinear solver.
Unfortunately, there is not currently a way to work out \emph{a priori} what
good choices of $\alpha$ and $q$ will be. In the situations where deflation was
used we have aimed to maximize the number of solutions obtained by scanning
through the $(\alpha,q)$-parameter space. 
%
To do this, whenever deflation is used in this work, we fix $q=1.5$ and try
many different $\alpha$ values in the range $(0, 1]$. (Whilst it would be more
comprehensive to scan through a range of exponents as well, this is much more time
consuming and was found to be a comparatively less effective way to locate
additional solutions.) The exact values of the shifts used are not as important
as the need to cover a range encompassing different orders of magnitude. We begin
deflation with an initial $\alpha_0$, typically in the range $[0.5,1]$ and find
successive solutions until the nonlinear solver fails (e.g.\ due to exceeding
the maximum allowed iterations). Each time a new solution is found, it is used
as the new initial condition for the solver, after applying a small perturbation
to ensure that the deflated operator is finite. 
After exhausting the solutions that we can find with the initial $\alpha_0$, we
continue, scanning through a geometric progression of shifts $\alpha_n :=
\tilde{r}\alpha_{n-1}$, until $\alpha_n < \alpha_{\mathrm{min}}$, whereupon
deflation is halted. For the systems considered in this paper,
$\tilde{r} = 2/3$ and
$\alpha_{\textrm{min}} = \num{5e-3}$ have been used.
At higher values of $\alpha$ the nonlinear solver stays near to the previously
deflated solutions since the non-deflated part of the system Jacobian is more
significant with respect to the deflated part. As $\alpha$ decreases, more
remote solutions become accessible, often at the expense of those with shapes
that are structurally close to the deflated ones. For small values of $\alpha$,
Newton's method may take very large steps that decrease the residual of
deflation operator, but not the residual of the original system. This can cause
numerical instabilities if it produces an intermediate guess which is highly
strained. To avoid this, we set an upper limit on the original system residual
which, if reached, causes the algorithm to reset the initial condition and move
on to the next $\alpha_n$.
Finally, we note that after a solution has been deflated, this does not prevent
Newton's method from taking steps towards it. In general, the solver is not
guaranteed to find a region where it will converge quadratically to a new
solution and can spend a long time approaching already-deflated results.  It is
not uncommon for the method to take more than 100 iterations to converge.  To
catch most of the solutions, we allow for a maximum of 300 iterations.

In addition to finding solutions with deflation, we were able to find a few
additional equilibria using parameter continuation. In this regard, the most
useful control parameter is $\kappa$.  Starting from an initial solution with
$\kappa = \kappa_0$, we gradually increment or decrement $\kappa$ until the
system adopts a qualitatively different shape. Then continuing $\kappa$
gradually in the reverse direction may produce a distinct
solution. The interpretation of this procedure is that the system passed a
bifurcation point, uncovering a new solution branch, which we can trace back to
$\kappa_0$.

Given a set of distinct solutions, it is desirable to determine which will be
preferred by the physical system. The potential energy $\Pi$ of the DE is given
by integrating the strain energy density over the whole body, minus the work
done by the body forces and tractions. This is
\begin{equation}
	\Pi(\vect{u}) = \int_{\Omega_0} W(\vect{u})\, dV - \int_{\Omega_0} \vect{b}_0 \cdot
	\vect{u}\, dV - \int_{\partial\Omega_0} \vect{\tau}_0 \cdot \vect{u}\, dA,
    \label{eq:elasticpotential}
\end{equation}
where $\vect{u}$ is a function that gives the displacement of a material point,
relative to its position in the undeformed configuration $\Omega_0$, and
$\vect{\tau}_0$ is the field of tractions on the domain boundary. We perform these
integrations numerically over the discretization mesh that we use to
solve Eq.~\eqref{eq:elastostatics}. This allows us to calculate the minimum energy
solution from the shapes found.
\section{Results}
\label{sec:results}
\subsection{Circular active region}
\begin{figure}[bp]
	\centering%
  \includegraphics[width=8.5cm]{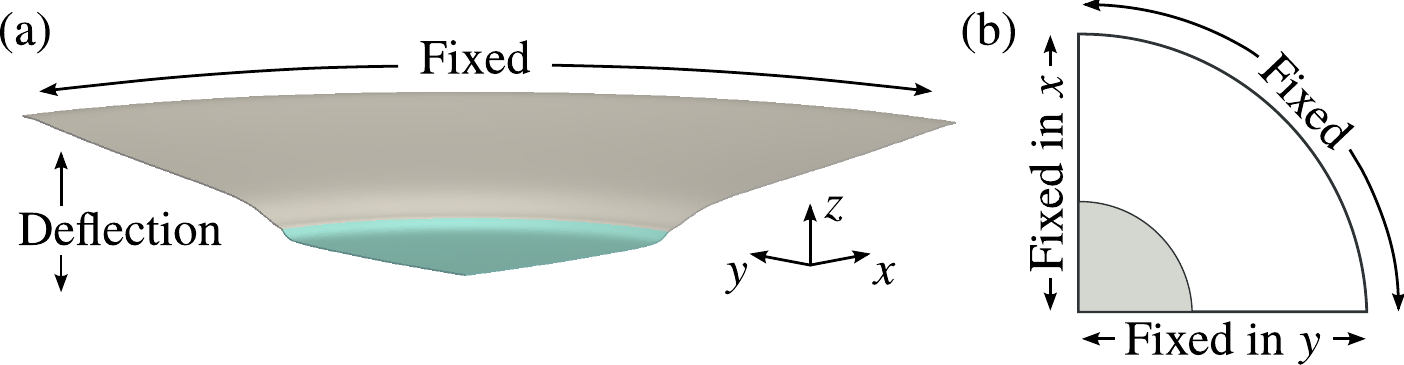}%
	\caption{
		(a)~Oblique view of a typical deformed configuration for a thin disc with
		circular active region, showing the overall deflection and
		localized buckling in the active region, which is shaded in
	        cyan. The inactive part is shaded in light gray and the fixed
	outer edge of the domain is indicated.
		Only a quarter of the geometry is
		simulated---the rest is accounted for via boundary conditions which
		enforce reflection symmetry in the planes $x=0$ and $y=0$. The diameter
		of the full geometry is $666\sfrac{2}{3}D_0$ and the diameter of
		the full active region is $l_0 = 266\sfrac{2}{3}D_0$. Other parameters
		are: $\kappa = 0.6$, $\tau_n = 0.5$, $\tau_t = \num{0.0225}$ and $\rho g
		= 0$.
		(b)~Schematic showing the boundary conditions for
		the circular disc.}%
	\label{fig:circularExamples}
\end{figure}

Before delving into the details of matching simulations with experiment, we
present a representative simulation of a buckled DE with applied normal and
tangential tractions. 
Figure~\ref{fig:circularExamples}(a) shows a solution for
a circular disc-shaped elastomer with a circular active region at the center. 
The plot is an oblique view of the deformed configuration, with the active 
region indicated.
To save computational effort, we solve the elastostatics equation for only a
quarter of the axisymmetric geometry.  Consequently, we see a cross-section of
the elastomer in the figure and may easily inspect the solution's out-of-plane
deflection. Starting from the outer edge, the profile slopes gently
downwards, before an abrupt transition at the edge of the active region where
the gradient becomes much steeper. In the bulk of the active region however,
the profile levels out and is close to flat.
Figure~\ref{fig:circularExamples}(b) indicates the boundary conditions used. The
outer arc of the disc is fixed in place with a Dirichlet condition. The other
two edges are free to move both in the radial direction and out-of-plane
($z$-direction), while their remaining degree of freedom is fixed. Solutions for
these boundary conditions correspond to solutions to the full problem with at
least reflection symmetry in the $x$ and $y$ directions.  (In practice our
circular-electrode solutions possess continuous rotational symmetry in the
$xy$-plane.)

In each of the following cases the geometry of the simulation is set such that
the dimensions of the finite element mesh equal those of the experiment. We set
our model parameters using a nondimensional approach, taking the undeformed
material thickness $D_0$ to be the natural length unit for the system. 
We choose $s_0$ such that the tangent force is applied over a width of at least
two (quadratic) finite elements. In all results, $10\leq s_0 / D_0 \leq 20$.
After fixing the geometry, there are five free parameters in the model: the
Mooney-Rivlin constants $c_1$ and $c_2$, the density $\rho$ and the tractions
$\tau_n$ and $\tau_t$. 
We are free to choose $c_1 = 1$, since it may be easily verified that
any solution to the governing equations satisfies those same equations after
rescaling each model parameter $k$ by $k\mapsto \mu k$ for any nonzero
constant $\mu$.
Moreover, we found that varying the ratio $c_1 / c_2$ had no noticeable effect
on the shape of our solutions in any of the contexts studied herein. Hence, we
set $c_1 = c_2 = 1$ throughout.  The redundancy of the $c_2$ parameter suggests
that, at least for the range and type of strains that we consider, a
Neo-Hookean constitutive law ($c_2 = 0$) may be sufficient to model the
elastomer well.

\begin{figure}[htbp]
	\centering%
  \includegraphics{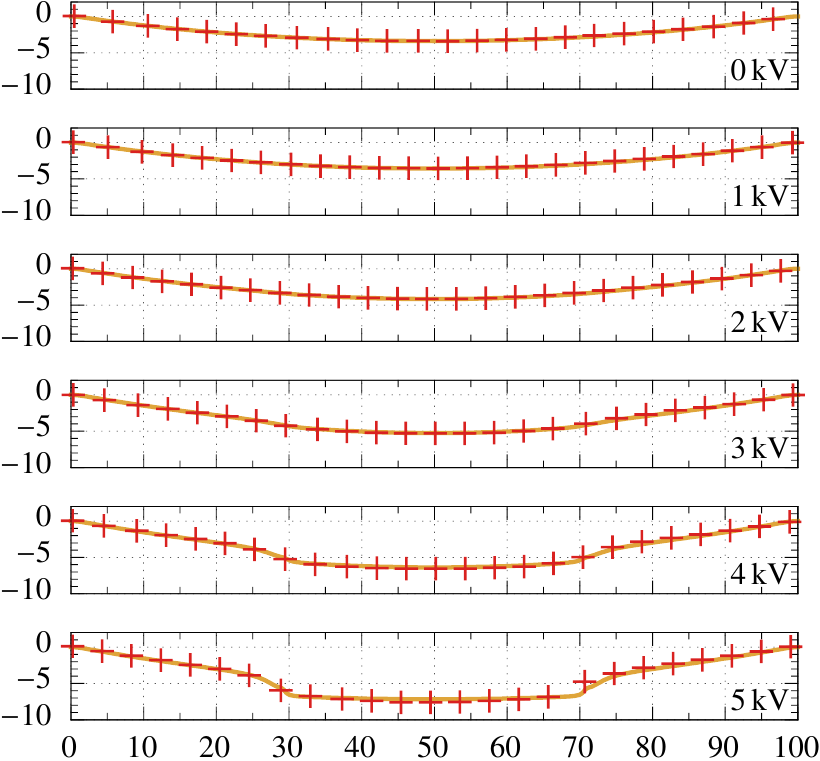}%
	\caption[Comparison of experimentally measured deflections with simulation
    profiles for a succession of increasing voltages.]
    {
		Comparison of experimentally measured deflections with simulation
		profiles for a succession of increasing voltages:
		\SIlist{0;1;2;3;4;5}{\kV}.  The aspect ratio is $1:1$. Red crosses
		indicate experimental data points.  The experimental geometry used was a
		thin disc, diameter $\SI{100}{\mm}$, thickness $D_0 = \SI{0.15}{\mm}$,
		with centered circular electrodes, $l_0 = \SI{40}{\mm}$. Yellow lines are
		midlines through a model simulation with corresponding geometric
		parameters and $\kappa = 0.6$ in each case.  
		The gravitational body force is $\rho g = \num{3.6e-4}$
		throughout. Applied tractions across the different voltages are (to 4
		significant figures): $\tau_n = \tau_t = 0$ for \SI{0}{\kV}; $\tau_n =
		\num{0.01857}$, $\tau_t = \num{8.538e-4}$ for \SI{1}{\kV}; $\tau_n =
		\num{0.07633}$, $\tau_t = \num{3.435e-3}$ for \SI{2}{\kV}; $\tau_n =
		\num{0.18}$, $\tau_t = \num{8.1e-3}$ for \SI{3}{\kV}; $\tau_n =
		\num{0.347}$, $\tau_t = \num{0.01561}$ for \SI{4}{\kV}; $\tau_n =
		\num{0.6176}$, $\tau_t = \num{0.02779}$ for \SI{5}{\kV}.
	}%
	\label{fig:fitmultiplevoltages}
\end{figure}

Figure~\ref{fig:fitmultiplevoltages} shows comparison between simulation and
experiment for six different applied voltages.
Each plot shows the midline of a numerical solution restricted to the $y = 0$
plane, together with points of experimentally measured deflection. The
experimental data covers the full diameter of the elastomer. Therefore, the
simulation midline in this case is mirrored across the axis of symmetry in the
plots.  
When taking measurements in the experiment, the deflection of the surface means
that the laser does not travel exactly through the elastomer diameter.
Consequently, the experimental data does not extend fully to the edges of the
simulation domain. In order for the experiment and model geometries to match (in
particular the electrode radii), it is necessary to apply a correction.
Therefore, we adjust the horizontal scale of the experimental points by a small
amount (4.2\%), chosen such that the edges of the laser trajectory match those
of the simulation domain.

The procedure for fitting the model parameters is as follows. First, the
profile of the elastomer with no applied voltage is measured. In this case,
there is only one free model parameter---the material density---which is
adjusted in the simulations until the amplitude at the center
matches the experiment. Next, voltage is applied in the experiment to produce
significant additional strain in the elastomer and the profile is measured
again, in this case at \SI{3}{\kV}. The nontrivial shape adopted by the data
points allows us to fit both $\tau_n$ and $\tau_t$ concurrently and thereby
determine $\kappa$ [Eq.~\eqref{eq:kappa}]. This is because the amplitude of the
active region deflection and the overall shape at the electrode boundary are
effectively independent of one another in the model.
The amplitude of deflection 
corresponds roughly to the total applied traction and while the shape at the
electrode boundary is determined by the ratio $\tau_t / \tau_n$.
We will return to this point shortly. 
%
%
After making an initial guess of their approximate magnitudes and ratio,
$\tau_n$ and $\tau_t$ are incrementally increased or decreased (in concert)
until the solution amplitude matches the experiment. Next, to match the profile
shape, $\tau_t$ is incremented or decremented.  
  Small discretionary adjustments to the tractions are then made to
improve agreement further.  From this point on, both $\rho$ and $\tau_t /
\tau_n$ are considered to be fixed.

We know from Eq.~\eqref{eq:classicPressure} that the normal pressure $p$ is
proportional to $(V/D)^2$.  The two fitted results at \SIlist{0;3}{\kV} uniquely
determine the coefficient of proportionality. 
However, since $p$ depends on the deformed thickness $D$, the amount of normal
pressure for a given voltage is coupled to the solution. Furthermore, we model
$p$ with the applied traction $\tau_n$ in the Lagrangian frame and the pressure
that a given $\tau_n$ corresponds to in the deformed body also depends on $D$.
Specifically, one can show that $\tau_n \propto p / D \propto V^2/D^3$.
Therefore, for the remaining voltages in Fig.~\ref{fig:fitmultiplevoltages}, we
determine $\tau_n$ using an iterative procedure. Each step enforces the
proportionality condition using the deformed thickness (taken at the center
point) of the previous iteration. A similar approach was used in
Ref.~[\onlinecite{Wissler2007a}].
Successive iterations converge rapidly to a normal pressure that scales
correctly with the electric field in the experiment. The applied tractions used
in Fig.~\ref{fig:fitmultiplevoltages} all obey the correct scaling relation dictated
by Eq.~\eqref{eq:classicPressure} to within $1\%$ relative error.  Throughout this
procedure, $\tau_t$ is selected such that $\tau_t/\tau_n$ (and thus $\kappa$)
stays the same.

The profile shapes obtained this way agree extremely well across all the plots,
even though the parameters were only fitted using the $\SI{0}{\kV}$ and
$\SI{3}{\kV}$ cases.  For voltages greater than or equal to
\SI{4}{\kV} there are very small discrepancies which may, for instance, be due to the
constitutive law used, or the simplified treatment of the forces acting on the
elastomer in our model.  Nevertheless, even at these higher strains agreement
between the model and experiment is good.

\begin{figure}[t]
\begin{centering}
  \includegraphics{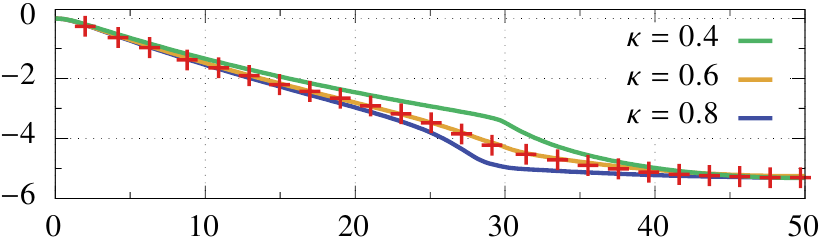}%
\caption{
	Effect of tangential shear on the shape of model profiles. The vertical
	axis has been scaled by a factor of 2 to show the variation between the
	profiles more clearly. Red crosses are data points from the \SI{3}{\kV}
	experiment in Fig.~\ref{fig:fitmultiplevoltages}.  
	Green, yellow and blue lines are model
	results with $\kappa = 0.4$, $0.6$ and $0.8$ respectively. In each case, the
	total traction was chosen so that the model profile matched the experimental
	deflection in the center, at $x = 50$. The tractions were as follows:
	$\kappa = 0.4$ used $\tau_n = \num{0.156}$, $\tau_t = \num{4.68e-3}$,
	$\kappa = 0.6$ used $\tau_n = 0.18$, $\tau_t = \num{8.1e-3}$ and $\kappa =
	0.8$ used $\tau_n = 0.22$, $\tau_t = \num{0.0132}$. All other model
	parameters match those from Fig.~\ref{fig:fitmultiplevoltages}.}%
\label{fig:tangentforceeffect}
\end{centering}
\end{figure}

For the tractions used in this particular case with a circular active region
centered inside a disc, $\tau_t/\tau_n = 0.045$. Taking into account the
geometric parameters, this corresponds to $\kappa = 0.6$. 
%
Since $\tau_t$ is so small compared with $\tau_n$, one may wonder whether the
tangential forces in the model may be neglected altogether. However, despite its
magnitude, slight changes in $\tau_t$ can have a marked effect on solutions.
Indeed, we find that $\kappa = 0.6$ fits the experimental data better than
either $\kappa = 0.58$ or $\kappa = 0.62$, though the differences between
model profiles are subtle at this level.
Figure~\ref{fig:tangentforceeffect} demonstrates the much more significant
effect of changing $\kappa$ by $\pm 0.2$.
Here, the \SI{3}{\kV} experimental data from Fig.~\ref{fig:fitmultiplevoltages} are
replotted alongside three model
profiles with $\kappa = 0.4$, $0.6$
and $0.8$. As $\kappa$ increases, the proportion of tangential force increases.
This has two main effects. Increased tension at the edges causes the active
region to flatten out and stretch. This in turn modifies the shape at the
electrode boundary. Both the $\kappa = 0.4$ and $\kappa = 0.8$ profiles feature
an abrupt change of gradient near the active region edge. Only $\kappa = 0.6$
features the smooth transition from inactive to active region that matches the
experiment. Thus, for all further
results, we use $\kappa = 0.6$ unless
otherwise stated. 

\subsection{Annular active region}
\begin{figure}[tbp]
    \centering%
    \includegraphics[width=5.5cm]{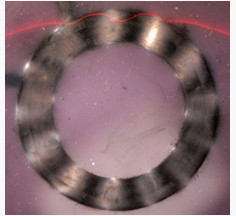}%
	\caption{
		Overhead view of an elastomer experiment with an annular active
		region, whose geometry corresponds to the simulation in
		Figs.~\ref{fig:annularExamples}(b) and (c). The inner
		radius of the annulus is $r_0 = \SI{17}{\mm}$, the outer radius is $R_0
		= \SI{25}{\mm}$ and $D_0 = \SI{0.15}{\mm}$. The applied voltage is
		\SI{3}{\kV}. Azimuthal ripples are visible on the electrode; their
		undulation is highlighted by directing a laser across the surface.}%
    \label{fig:annularExperimentPicture}%
\end{figure}

Another system of experimental interest is shown in
Fig.~\ref{fig:annularExperimentPicture}.
In this case, the active region is annular.
For sufficiently high applied voltage, this DE readily buckles to produce
azimuthal ripples in the active region. Wavelengths measured from the experiment
are robust over a range of voltage (\SIrange{3}{5}{\kV}) and depend principally
on the width of the annulus. In particular, increasing the applied voltage from
the onset of this instability only acts to increase the overall deflection of
the DE and amplitude
of its ripples. These ripples in the active region are distinct
from the much smaller wavelength wrinkles that result from a pull-in
instability.\cite{Plante2006}

\begin{figure}[htbp]
    \centering%
    \includegraphics[width=8.5cm]{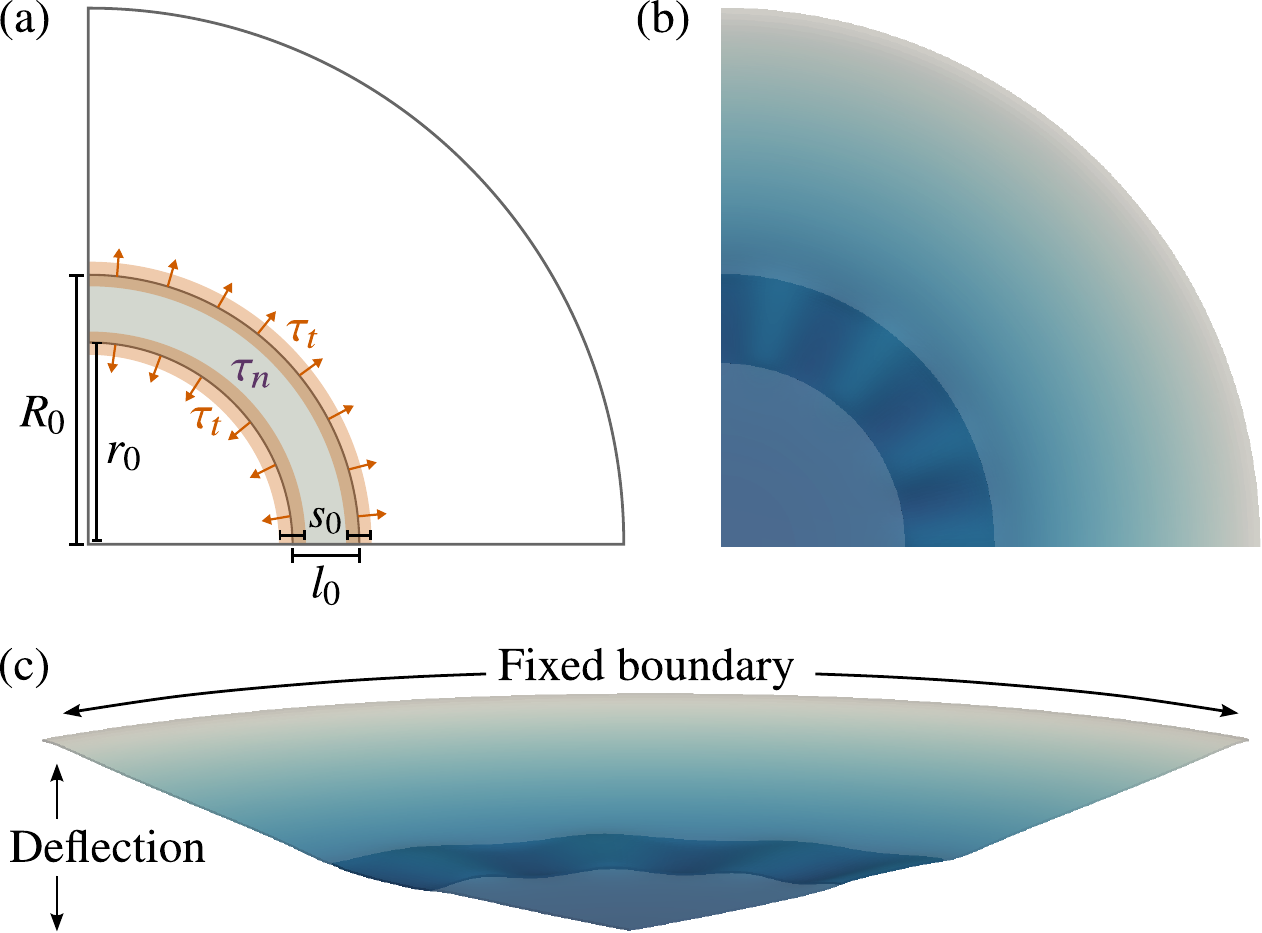}%
	\caption{
		(a) Diagram showing the top/bottom
	surface of the model setup for a circular disc with annular active region.
	Compressive normal pressure $\tau_n$ is applied into the page across the
	shaded gray area.  Tangential surface tractions $\tau_t$ are applied at
	both boundaries of the active region in the two orange areas shown. 
	Important length scales are labeled: the inner radius $r_0$ and outer
	radius $R_0$ of the active region annulus, its width $l_0$ and the
	width $s_0$ over which the tangential traction is applied.
	(b) Example result from the setup depicted in part~(a). The blue coloration
	indicates deformation in the negative $z$-direction. Deeper blue means that
	a point is displaced further below its original position in the flat
	reference configuration. The active region is indicated as an area of
	comparatively darker shading.  The geometry is set to match an experiment
	with $r_0 = \SI{17}{\mm}$, $R_0 = \SI{25}{\mm}$, $D_0 = \SI{0.15}{\mm}$ and
	diameter $\SI{100}{\mm}$.  Other model parameters are: $\kappa = 0.6$
	and $\rho g = \num{3.6e-4}$.  (c) Oblique view of the result in part~(b)
showing the overall deflection of the DE and the azimuthal ripples in the active region.}
	\label{fig:annularExamples} 
\end{figure}

In the annular case, the active region has two
edges. Consequently, there is an additional tangential fringing effect pointing
radially inward. A diagram of the simulation domain and imposed tractions is
shown in Fig.~\ref{fig:annularExamples}(a).
The two fringing regions are both modeled with width $s_0$ centered at the
inner and outer radii of the electrode annulus, labeled $r_0$ and $R_0$
respectively. The characteristic length scale of the active region $l_0$ in this
case refers to the width of the annulus.

We use the same boundary conditions as for the
circular disc [see Fig.~\ref{fig:circularExamples}(b)], simulating only a
quarter
segment of the whole system in order to save computational cost. However, in
this case, our buckled DEs do not possess continuous rotational symmetry.
Therefore, it is important to note that these conditions place
constraints on the range of admissible wavelengths. 
In cases where the wavelengths are particularly large, we increase our domain 
size to half a disc, ensuring that the simulation can always fit many 
ripples within the given domain. 

Figures~\ref{fig:annularExamples}(b) and (c) show overhead and oblique views of
a simulated result for an annulus of width $53\sfrac{1}{3}D_0$. The dimensions
of this simulation correspond to the experiment photograph in
Fig.~\ref{fig:annularExperimentPicture}. 
From visual inspection one sees a qualitative agreement between the experiment
and simulation, both in the overall deformation profile and the character of the
ripples.

As mentioned above, there may be many distinct
solutions to the elastostatics equation [Eq.~\eqref{eq:elastostatics}] that are
not related by symmetry. 
Indeed, for this system it is possible to find solutions with different
azimuthal wavelengths.
The wavelength selected by the physical system would typically be the one which
minimizes the
energy, given in Eq.~\eqref{eq:elasticpotential}. This
is not generally the solution first discovered by our nonlinear solver. To
overcome this problem, we use the deflation method, described in
Sec.~\ref{sec:deflation}, to find as many different solutions as we can.
The result pictured in Figs.~\ref{fig:annularExamples}(b) and (c)
is the minimum energy solution of four different equilibrium configurations
computed by this technique. Likewise, the annular active region results below
are all energy minima from sets of deflated solutions. However, deflation does not
guarantee that every solution will be found. To increase our confidence that
these results are close the global minima, we can compare their azimuthal
wavelengths with measurements from the experiment. 

\begin{figure}[htbp]
	\centering%
  \includegraphics[width=8.5cm]{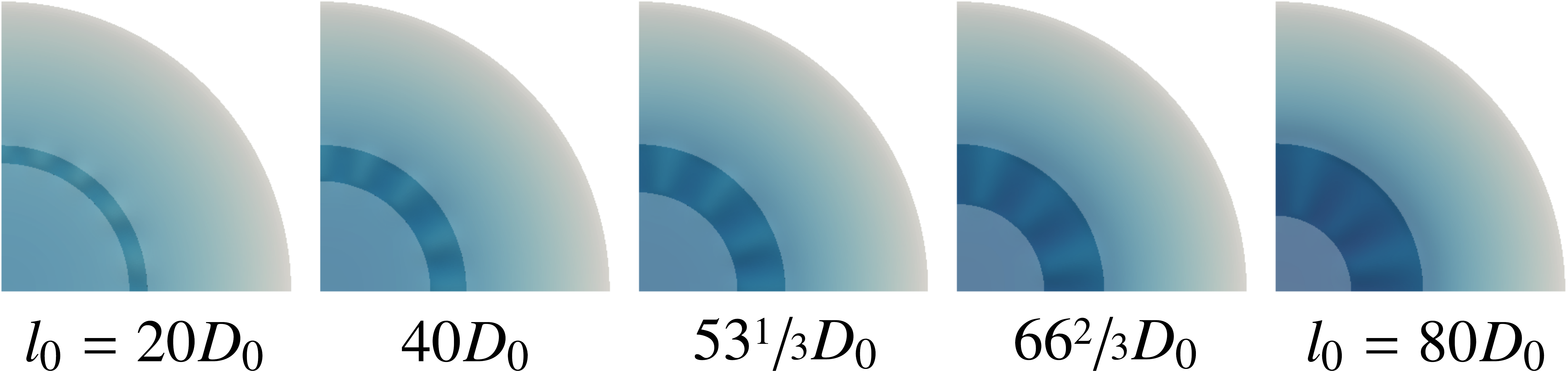}%
	\caption{%
		Deformed configurations for a circular disc with annular active regions
		of different widths $l_0$. Each is the solution found with the lowest
                energy after deflation. As $l_0$ increases, so does the
		wavelength of ripples in the active region. The extent of the active
    	        region in each case is indicated with darker shading. Model
	        parameters are the same as in Fig.~\ref{fig:annularExamples}, save $r_0$
                which was adjusted for each $l_0$ as indicated.}%
	\label{fig:annularDifferentl0}
\end{figure}

Figure~\ref{fig:annularDifferentl0} shows simulations with various annular active
region widths. 
\begin{figure}[htbp]
    \centering%
    \includegraphics[width=\columnwidth]{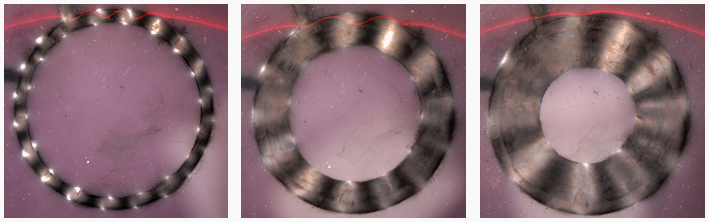}%
	\caption{
		A selection of annular experiments with $R_0 = \SI{25}{\mm}$
		and different $l_0$. The applied voltage is \SI{3}{\kV}.
		From left to right: $l_0 = 3, 8$ and \SI{12}{\mm}. 
		The geometries correspond to the first, third and fifth simulations
		in Fig.~\ref{fig:annularDifferentl0} respectively.}
    \label{fig:annularExperimentDifferentl0}
\end{figure}
One sees that as $l_0$ increases, the wavenumber observed across the quarter
segment decreases. This is observed in experiment: in
Fig.~\ref{fig:annularExperimentDifferentl0} we show overhead pictures of the experiment
with different annular widths. 
These correspond to the simulated geometries with $l_0=20D_0$,
$53\sfrac{1}{3}D_0$ and $80D_0$ and may be compared directly with the pictures
in Fig.~\ref{fig:annularDifferentl0}. Qualitatively there is good agreement between
the two sets of images.

\begin{figure}[bp]
    \centering%
    \includegraphics{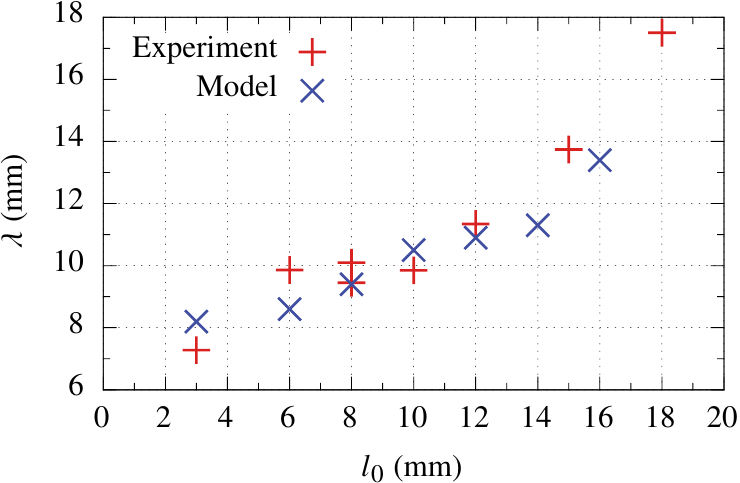}%
    \caption{
	Width $l_0$ of annular active region versus observed ripple
        wavelength $\lambda$ for the experiment and model simulations. The
		experiment parameters were $D_0 = \SI{0.15}{\mm}$ with $R_0$ fixed at
		$\SI{25}{\mm}$ and various $r_0$ between $7$ and $\SI{22}{\mm}$. The
        applied voltage was \SI{3}{\kV}.
	}%
    \label{fig:l0_vs_lambda}
\end{figure}

In Fig.~\ref{fig:l0_vs_lambda} we plot both experimental and simulated ripple
wavelengths against $l_0$ and see more clearly the quantitative agreement
between the two.
Wavelengths are calculated in both cases by dividing the circumference of the
circle of radius $r_0 + \tfrac{1}{2}l_0$ by the observed wavenumber.  The
results with $l_0 = 12$, $14$ and $16$ were obtained with a half-disc simulation
domain.  There is a degree of uncertainty associated with measuring these data
points experimentally.  
Nevertheless, the model does a good job of matching the smaller reported
wavelengths in the physical system.

\begin{figure}[tbp]
    \centering%
    \includegraphics[width=\columnwidth]{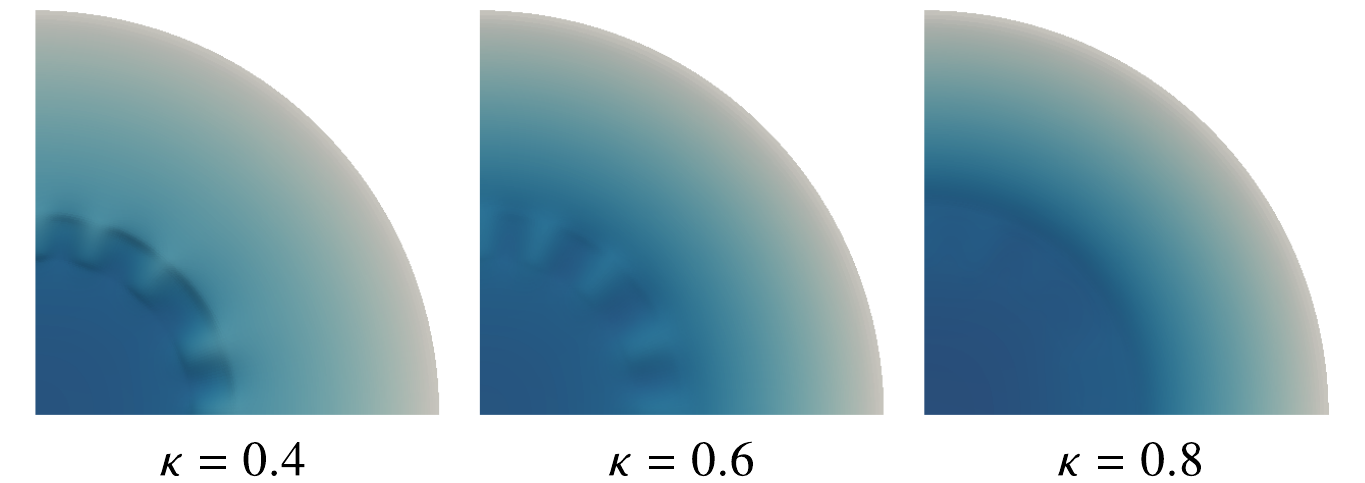}%
    \caption{
	    Effect of tangential shear on annular active region
	simulations. All model parameters match those in
	Fig.~\ref{fig:annularExamples}, aside from $r_0$ which was set to give
	$l_0=40D_0$ and $\tau_t$ which was varied for different $\kappa$ as
	labeled.
	}%
    \label{fig:annular_different_kappa}
\end{figure}

Finally, we verify that $0.6$ is indeed a good choice for $\kappa$, as it was
in the case of circular electrodes.  Figure~\ref{fig:annular_different_kappa}
plots simulations of the $l_0=40D_0$ case for $\kappa = 0.4, 0.6$ and $0.8$.
To produce these solutions, we fix $\tau_n = 0.32$ and vary $\tau_t$ to achieve
the desired $\kappa$. Decreasing the amount of tangential shear to $\kappa =
0.4$ causes the edges of the active region to crease slightly and the spacing
between ripples becomes uneven. Increasing to $\kappa = 0.8$ flattens the active
region and the ripples disappear. In both cases, the effect on the radial
deflection profile is similar to Fig.~\ref{fig:tangentforceeffect}.

\subsection{Rectangular active region}
\begin{figure}[htbp]
	\centering%
  \includegraphics[width=8.5cm]{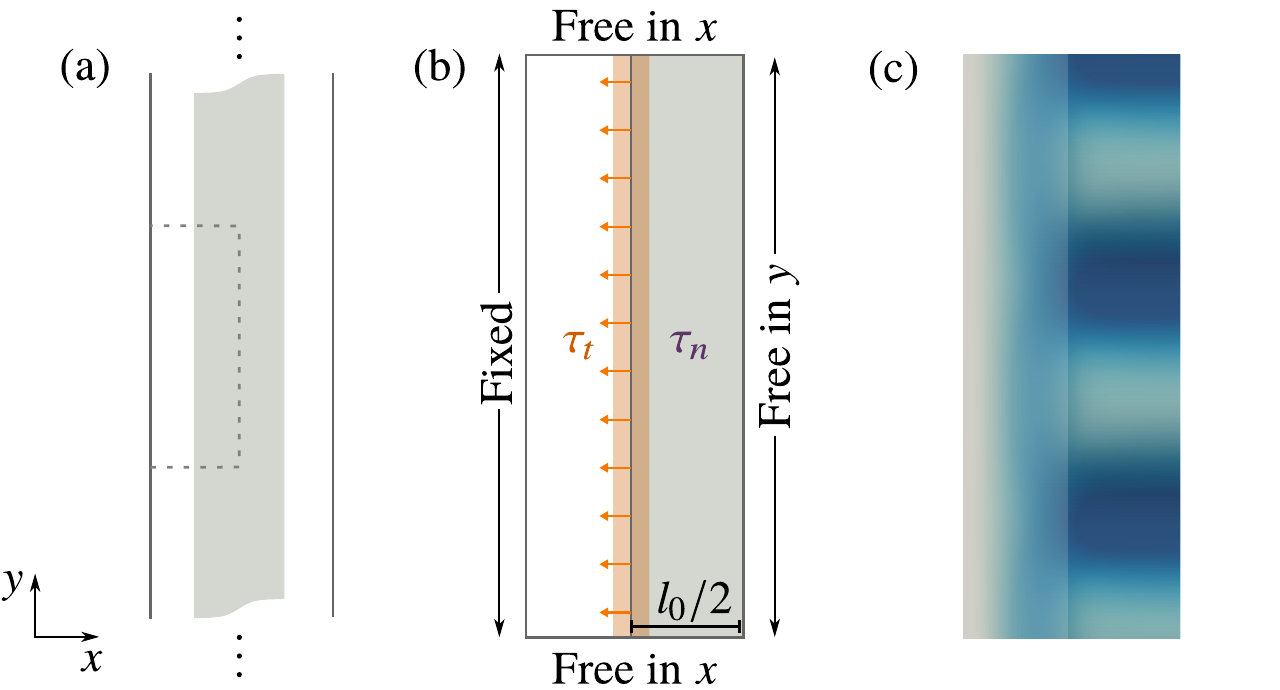}%
	\caption{
		(a) Top-down picture of a rectangular strip DE, with active
		region shaded in gray. Dots indicate that the elastomer extends
		far in its lengthwise extent. The dashed gray rectangle shows the
		region simulated in our computations, which take advantage of
		symmetries described below.
		(b) Diagram of the boundary conditions for simulation.
		Periodic symmetry is enforced at the top and bottom edges. The
		other two edges implement reflective symmetry in the axis along the
		right-hand side. See the text for details. Surface tractions $\tau_n$
		and $\tau_t$ are applied in the gray and orange regions respectively, as
		indicated.
	The definition of the characteristic width $l_0$ for this active region is
	as labeled. It covers half the simulated domain.
	(c) An example deformed 
	configuration. The active region is indicated with an area of darker
	shading. As in the annular case, ripples are present. These follow the 
	direction of the strip's longer dimension.
	The model parameters are: $\kappa = 0.6$, $\tau_n =
	0.37$, $\tau_t = \num{0.0222}$, $\rho g = 0$ and the dimensions are height
	$H_0 = 250D_0$ and width $L_0/2 = 92\sfrac{1}{2}D_0$.
	}%
	\label{fig:rectangleExamples}
\end{figure}

A third simple, but important configuration is a long rectangular elastomer with
a rectangular active region, as illustrated in
Fig.~\ref{fig:rectangleExamples}(a).  Provided that the length of the rectangle
is sufficiently greater than its width, this system also readily buckles to
produce ripples along its length.  This was previously noted by Pelrine \emph{et
al.}\cite{Pelrine2000} Similar ripples in a prestretched DE were also
observed by D\'{i}az-Calleja \emph{et al.}\cite{Diaz2013} Our own experimental
investigations, while not extensive, indicate that the ripple wavelengths are
approximately equal to those observed in an annular active region of the same
width.


%
A schematic of our model setup is shown in
Fig.~\ref{fig:rectangleExamples}(b), alongside a representative result in
Fig.~\ref{fig:rectangleExamples}(c).
We consider a rectangular geometry with its width aligned with the $x$-axis and
its length aligned with $y$. We assume that deformations are symmetric about the
midline in $x$ and so only simulate half the full width. Furthermore we, set the
active region to the full length of the domain. This effectively mimics a
portion of longer elastomer far from any physical boundaries in $y$.  All
together this means that the active region extends to the edges of the
computational domain on three sides.


The boundary conditions are as follows.  The two short edges of the domain are
free to move in the $x$ and $z$ directions only. Fixing them in $y$ enforces
periodic symmetry.\footnote{Note that since we do not require each end to deform
to the same height in the $z$-direction, the enforced periodicity is equal
to twice the length of the simulation domain. Fully periodic solutions may be
obtained by a reflection at either end.} No tangential force is applied in this
direction.
One of the two long edges is held fixed, corresponding a frame holding the
elastomer. The other long edge is free to move in the $y$ and
$z$-directions, corresponding to the reflection symmetry about the midline. 
%
%
The dimensions of the simulation domain are height $H_0 = 250D_0$ and width
$L_0/2 = 92\sfrac{1}{2}D_0$. The width $l_0/2$ of the simulated active region
covers half the domain.

Similar to the case of an annular active region,
the finite extent of the domain means that some long wavelengths are
inaccessible.
Guided by the results in the annular case and intuition from experiments with
rectangular electrodes, we believe that the domain dimensions chosen are
sufficient to capture any important solutions. 

\begin{figure*}[htb]
    \centering
    \includegraphics[width=1.75\columnwidth]{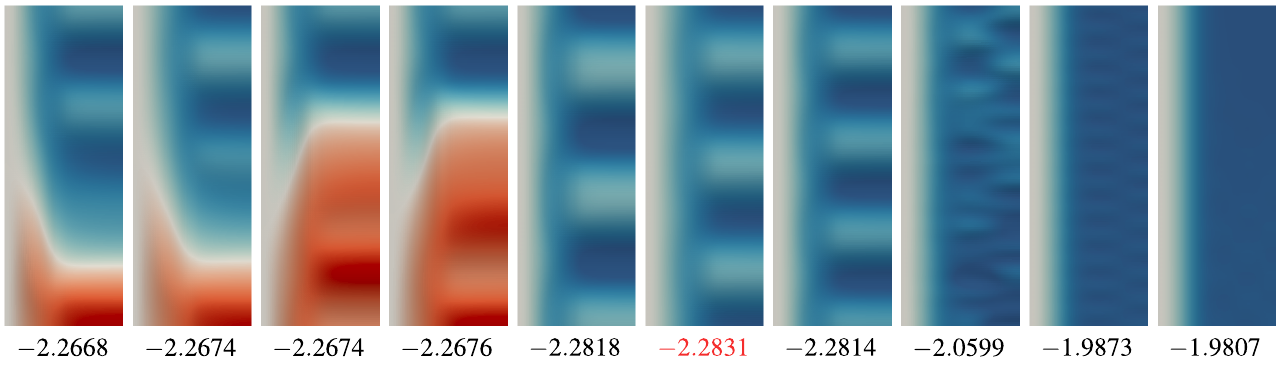}%
    \caption{
	Deflated solutions for a long rectangular strip. $\kappa =
        0.6$. The blue and red coloration indicate deformation in the
        $z$-plane. Darker red (blue) means that a point is displaced further
        above (below) its original position in the flat reference state. Beneath
        each solution the 
        energy computed using
        Eq.~\eqref{eq:elasticpotential} is printed. 
	Solutions that are equivalent under symmetry to the ones shown have been
	omitted. 
	}
    \label{fig:rectdeflfton}
\end{figure*}

We were able to obtain many solutions via deflation for this geometry. These are
shown in Fig.~\ref{fig:rectdeflfton}, with their corresponding
energies printed underneath. 
In this case, a variety of interesting
solutions can be found.
To this end, we omitted the gravitational body force
from the model. This encourages the DE to buckle up, as well as down and enables
us to find more solutions. 
In the center of the figure are solutions with regular ripples, analogous to
those seen in the annular active region. 
To the left, there are four
solutions composed of a large wavelength mode and smaller ripples.  To the
right are solutions with higher frequency ripples: one with regular ripples,
another with irregular ripples and one with a smooth, mostly flat active
region.
For each shape shown,
reflections in the planes $y=H_0/2$ and $z=D_0/2$ give solutions that are
equivalent
under the symmetries of the problem. These have been omitted
from Fig.~\ref{fig:rectdeflfton}.
The final two solutions to the right were found using the parameter continuation
technique described in Sec.~\ref{sec:deflation}.
%
The highlighted entry is the minimum energy solution. It is important to
note that this was not the first solution to be found by the nonlinear solver.
In fact, prior to using deflation, the only configuration accessible was
	the leftmost solution in Fig.~\ref{fig:rectdeflfton}, which does not even
	agree qualitatively with the minimum.
Therefore, in this case it was essential to use deflation (or some alternative
method) to find multiple solutions and thereby identify the correct equilibrium
DE shape. We note that finding a higher-energy solution at lower wavenumber
gives us reason to believe that increasing the domain length will not produce a
lower energy minimum.
Finally, preliminary experimental investigations
indicate that the minimum energy numerical solution captures both the wavelength and
amplitude of rippling for a rectangular strip.
\section{Discussion}
We have presented a simplified numerical model for capturing the shape of
buckled DE. The electrostatic forces acting on the dielectric are input as
boundary conditions to the nonlinear elastostatics equation. We have proposed
that the aggregate effect of the applied electric field on the elastomer can be
modeled as a normal pressure, due to the attraction between the electrodes, plus
a small tangential traction meant to capture the effect of the fringing field at
edges of the
active regions.  The resulting boundary conditions are easily implemented and
while they represent a simplification of the underlying physics, they are
nonetheless able to produce close fits to experimental data.  

The magnitude of the fringing force, relative to the effective pressure is
captured by our model in a dimensionless constant $\kappa$.  By tuning
$\kappa$ to produce solutions best matching experimental deformation profiles,
we have found that $\kappa = 0.6$. This value proved robust across different
applied voltages and different shapes of active regions.  
%
The impact that the tangential traction has on solutions is significant, despite
its small magnitude. If the effect is left out of the model ($\kappa = 0$), we
are unable to obtain deformation profiles that are even qualitatively correct.

We have computed deformed solutions for a variety of active regions---circular,
annular and rectangular. For the circular and annular cases we have
quantitatively compared numerical solutions with experimental observations.  In
the case of an annular active region, we observe that the elastomer buckles to
produce azimuthal ripples, which are localized in the vicinity of the electrodes.
Their wavelength increases in proportion with the width of the annulus. This
trend is captured well by our model which produces solutions in qualitative
agreement with the experiment.

Our approach is quite generic and could be used for a variety of
elastomer geometries. 
Furthermore, the model
is, in principle, amenable to arbitrary active region shapes, though some care would
need to be taken at any nonsmooth features such as corners. Prestretch may be
applied by adjusting the dimensions of the unstrained reference configuration
$\Omega_0$, relative to the imposed Dirichlet (fixed-displacement) boundary
conditions at the domain edges. 
While we would not necessarily expect our model to
be predictive in high-strain regimes, it may be useful in some circumstances.
For instance, one effect of prestretch is to nonlinearly increase the buckling
threshold.\cite{Bense2017} Although the basic mechanism is clear, the nonlinear
dependence of the threshold is not currently understood. Since this phenomenon
occurs at low prestretch, a careful application of our model might capture it.

Finally, a key aspect of our study is the computation of multiple solutions.
We demonstrate that non-uniqueness of equilibria must be 
considered whenever model configurations are generated---a fact that has
implications for any study of patterns in nonlinear elasticity.
In computing a single solution
to the elastostatics equation [Eq.~\eqref{eq:elastostatics} or
Eq.~\eqref{eq:elastostaticsRef}], one cannot guarantee that it corresponds to the
equilibrium shape with the lowest possible potential energy. Indeed, we
observe that for a given set of model parameters, the first solution located by
our nonlinear solver (damped Newton's method) is typically not energetically
favorable. Consequently, it
is desirable to find many different solutions and work out which is favored
by the system, either by computing their potential energies
via Eq.~\eqref{eq:elasticpotential}, comparing with experimental data, or using
some other physical argument.
Deflation is one such technique that can be used to find multiple
solutions.\cite{Farrell2015} For the annular active region, this was used to
find the lowest energy azimuthal wavelength, which was subsequently compared
with experimental observations in Fig.~\ref{fig:l0_vs_lambda}. 
Almost all of the model wavelengths reported in our paper come from
solutions that were only found after applying the deflation method.
In the analogous case of a long rectangular strip, the first solution
that we computed does not even qualitatively resemble the minimum. By computing
multiple rectangular solutions, we identified
many interesting deformation patterns including ripples with different
wavelengths,
wrinkles and creases.
\begin{acknowledgments}
We are grateful to B. Roman and J. Bico for introducing us to this
system and for productive discussions. JL acknowledges support from an EPSRC
Doctoral Prize Fellowship, Grant No.\ EP/N509619/1.
\end{acknowledgments}

\end{document}